\documentclass{iopart}

\usepackage{amssymb}
\usepackage{amsthm}

 \usepackage{epsfig}

\newcommand{\sect}[1]{\setcounter{equation}{0}\section{#1}}

\def\be{\begin{equation}}
\def\ee{\end{equation}}
\def\bea{\begin{eqnarray}}
\def\eea{\end{eqnarray}}

\def\p{\partial}
\def\a{\alpha}
\def\b{\beta}

\def\g{\gamma}

\def\l{\lambda}

\def\L{\Lambda}
\def\vfi{\varphi}
\def\wt{\widetilde}
\def\ctU{\widetilde{\cal U}}
\def\cU{{\cal U}}
\def\bU{{\Bbb U}}
\def\cV{{\cal V}}

\def\cvV{{\overline\cV}}
\def\cuU{{\overline\cU}}

\newcommand{\Sc}{Schr\"odinger }
\newcommand{\jc}{Jaynes-Cummings }

\begin{document}

 \large

\title {Darboux transformations of the Jaynes-Cummings Hamiltonian}

\author{
 Boris F Samsonov$^{\dag}$${}^\S$\ and
 Javier Negro$^{\dag}$
}

\address{\dag\ Departamento de F\'{\i}sica Te\'orica, Universidad de
Valladolid,  47005 Valladolid, Spain}

\address{\S \, Physics Department of Tomsk State
 University, 634050 Tomsk, Russia}

\ead{\mailto{samsonov@phys.tsu.ru},
\mailto{jnegro@fta.uva.es}
 }

\begin{abstract}
\baselineskip=16pt

\noindent
A detailed analysis of matrix Darboux transformations
under the condition that the derivative of the
superpotential be self-adjoint is given.
As a consequence, a class of the symmetries associated to
Schr\"odinger matrix Hamiltonians is characterized. 
The applications are oriented
towards the
\jc eigenvalue problem, so that exactly solvable $2\times 2$ matrix
Hamiltonians of the
\jc type are obtained.
It is also established that the \jc Hamiltonian is a
quadratic function of a Dirac-type Hamiltonian.

\end{abstract}

 \submitto{\JPA}

\medskip
\medskip



\medskip
\medskip

\textbf{Corresponding Author}:

Javier Negro

Departamento de F\'{\i}sica Te\'orica (Facultad de Ciencias)

Universidad de Valladolid

47011 Valladolid

SPAIN

\medskip

Tel.: 34 983423040

E-mail: {\it jnegro@fta.uva.es\/}


\newpage

\sect{Introduction}

Eigenvalue problems with matrix Hamiltonians play essential role
in many physical applications. For instance, in scattering theory
of composite particles they appear as multichannel
Hamiltonians. In quantum optics they are used for describing
multilevel atoms interacting with a quantized radiation field. In
the simplest case of a two-level atom this is a $2\times 2$ matrix
Hamiltonian and if the rotating wave approximation is used the
well-known exactly solvable version of the
Jaynes-Cummings (JC) Hamiltonian is involved
(see e.g. \cite{Orczag,mpjc}). Sukumar and Buck \cite{sukumar}
presented a number of exactly solvable models of a two-level atom
coupled to single mode radiation. Here we are addressing the
question if this class of exactly solvable Hamiltonians can be
enlarged with the help of the well-known supersymmetry 
approach (or in other words, by means of Darboux transformations).

While analyzing this question we were aware that the usual
multichannel supersymmetric quantum mechanics was almost useless for
this purpose. The reason is that this procedure is developed
under  the additional condition that the superpotential matrix $W(x)$
be Hermitian
\cite{msusy}
but, as we will show below, this is not just the case of the \jc
Hamiltonian. In this respect Andrianov {\it et al} \cite{Andr} already
considered the case where this assumption could be replaced by
a weaker demand, namely, the derivative of the matrix $W(x)$ is
Hermitian, which is equivalent to
$W(x) - W^+(x) = C$,
where $C$ is a skew-hermitian constant matrix. However, they excluded
the possibility for $C$ to be proportional to the Pauli
matrix $\sigma_2$ (thus making impossible to apply their method to the
\jc Hamiltonian) and moreover, their approach was developed in detail
only for a particular case. 
So, the main aim
of this paper is to analyze carefully the problem of hermiticity
of the potential difference produced by general Darboux transformations
\cite{GV,SP} and then apply these results to the \jc eigenvalue
problem.

Another interesting point we would like to stress here is a new
character of the factorization property that we observe when dealing
properly with the hermiticity problem. To show this feature, first we
find the transformation operator $L$, defined according to Goncharenko
and Veselov \cite{GV} as an intertwiner of two matrix Hamiltonians:
$Lh_0=h_1L$.
Once it is known, we can find its
formally adjoint $L^+$ and compose it with $L$ which gives us
$L^+L = h_0 + S$,
where $S$, in contrast to the usual SUSY approach \cite{msusy}, is not
proportional to the identity matrix but, in general,
it is a nontrivial (matrix)
operator. Since $L^+$ takes part in the adjoint intertwining
relation,
$L^+h_1=h_0L^+$,
the superposition $L^+L$ is a symmetry
operator for $h_0$ meaning that actually $S$
is a symmetry operator too.
Such an appearance of symmetry operators was first noticed in 
\cite{Andr}
and treated as a {\it hidden symmetry} exhibiting within supersymmetry.
Unfortunately, because of the above mentioned restriction imposed on
$C$, the method of \cite{Andr} cannot be applied to get the hidden
symmetries for the \jc Hamiltonian, so that in this repect we will make
use of our more general set up.

The paper is organized as follows. In the next section,
to fix the notations, we briefly describe the solution of the
multiphoton \jc eigenvalue problem pointing out how one can find
``non-physical" solutions. Section 3 is devoted to adapting the known
technique of matrix Darboux transformations to \jc Hamiltonians. In
section 4 we give a careful analysis of general matrix Darboux
transformation leading to Hermitian potentials. In particular, we find
that they allow for a new factorization scheme responsible of
appearance of non-trivial
 symmetry operators. We indicate a way for establishing a
 one-to-one correspondence between the spaces of solutions, and
 finally construct an integral transformation operator.
 Section 5 establishes the link between the results of the
 previous section and the \jc eigenvalue problem. In Section 6 we
 explicitly construct new exactly solvable matrix Hamiltonians of the
 \jc type. Some conclusions are drown in the last section.

\sect{Jaynes-Cummings eigenvalue problem\label{JC}}

Our starting point is the matrix Hamiltonian
\be\label{hjc}
h^{(k)}=\left(
\begin{array}{cc}
N+\a & \b a^k \\ \b ^*(a^+)^k & N
\end{array}
\right).
\ee
describing a multiphoton interaction between a two-level atom and a
single mode radiation field \cite{mpjc,sukumar}.
Here $a$ and $a^+$ are the standard harmonic oscillator creation
and annihilation operators, $N=a^+a$ is the number operator,
$\a \in \Bbb R$ and $\b\in \Bbb C$ are parameters.
The real integer $k$ corresponds to the number of photons the atom
exchanges with the field. For the most interesting particular
case of the one photon exchange, $k=1$, we will use the special
notation $h_{JC}:=h^{(1)}$.
In what follows we need not only the known ``physical" eigenfunctions of
of $h_{JC}$
corresponding to its discrete spectrum eigenvalues
but also those solutions of the \Sc equation
which do not belong to the Hilbert space.
Therefore, below we give a short outline of how these solutions may
be obtained.

Let $\psi_n$ be the usual number operator eigenstates,
$N\psi_n=n\psi_n$,
with the properties
$a^k\psi_n=(n+1-k)_k^{1/2}\psi_{n-k}$,
$(a^+)^k\psi_n=(n+1)_k^{1/2}\psi_{n+k}$.
Here
$(x)_k$ denotes the standard Pochhammer symbol,
$(x)_k=\Gamma(x+k)/\Gamma(x)$.
The Hamiltonian (\ref{hjc}) can be diagonalized by a rotation inside
any
 two-dimensional space spanned by the vectors $(\psi_n,0)^t$ and
 $(0,\psi_{n+k})^t$. (The superscript ``t" meaning the
 transposition.)
 Indeed, if we write an eigenvector $\Psi_E$ of
 $h^{(k)}$, $h^{(k)}\Psi_E=E\Psi_E$, in the form
 $\Psi_E=(c_1\psi_n,c_2\psi_{n+k})^t$, then the eigenvalue problem is
 reduced to a simple system of two linear homogeneous equations for the
 coefficients $c_1$ and $c_2$, which gives us the spectrum
\be\label{en}
E=E_n^{1,2}=n+\textstyle{\frac 12}(k+\a )
\pm \sqrt{\textstyle{\frac 14}(\a -k)^2+|\b |^2(n+1)_k}
\ee
and the coefficients
\be\label{c1}
c_2=\frac{|\b |}{\b }\,\frac{E-n-\a}%
{\left[\, |\b |^2(n+1)_k+(n+\a -E)^2 \right]^{1/2}}\,,\quad
c_1=\sqrt{1-|c_2|^2}\,.
\ee
The set of eigenvectors thus determined,
 $\Psi_n=(c_1\psi_n,c_2\psi_{n+k})^t$, $n=0,1,\ldots$
 is not complete in the
 Hilbert space but the
 missing eigenvectors can readily be obtained by
noticing that the vectors $\Psi_j^0=(0, \psi_j)^t$ are
eigenvectors of $h^{(k)}$, corresponding to an equidistant part of
the spectrum $h^{(k)}\Psi_j=j\Psi_j$, $j=0,\ldots ,k-1$.

An interesting feature of the Hamiltonian
(\ref{hjc}) is that for $k>2$ the square root in (\ref{en})
dominates for large $n$ leading to the unbounded character of the spectrum both
from above and from below.
 From
this point of view this Hamiltonian is similar to a Dirac
Hamiltonian.

The  easiest way to get
``unphysical" solutions for $h_{JC}$
is to notice that in the above construction we needed just a couple of
states $\{\psi_{n},\psi_{n+k}\}$ related by the operators
$a^k,(a^+)^k$. We can take also these vectors from the unphysical
sector.

For instance, choosing them among the
basis vectors $\{\psi_{-n}\}_{n=1}^{\infty}$ of the
skew-hermitian representation of the oscillator algebra
where
$a^+\psi_{-n}=\sqrt{-n+1}\,\psi_{-n+1}$,
$a^-\psi_{-n}=\sqrt{-n}\,\psi_{-n-1}$, $a^+\psi_{-1}=0$ we will get
eigensolutions $\Psi=(c_1 \psi_{-n},c_2 \psi_{-n+k})$ corresponding to
an eigenvalue $E$. The expressions for $E$ and $c_1,c_2$ are given in
this case by the same formulae (\ref{en}) and (\ref{c1}),
where $n$ has to be simply replaced by $-n$. In the coordinate
representation where 
$\psi_n(x)=(\sqrt \pi n!2^n)^{-1/2}\exp (-x^2/2)H_n(x)$,
$H_n(x)$ being the Hermite polynomials, the negative eigenvectors
are realized as $\psi_{-n}(x) = \psi_{n-1}(i x)$.
Here
the eigenvalues can be complex according to the sign under the square
root  in (\ref{en}) that now may be negative depending on $n,k,\a$,
and $\b$. A particular set of nonphysical eigenvectors is given by
$(\psi_{-j},0)^t$,
$j=1,\dots,k-1$.

Even we can use a general solution $\xi_1(x)$ of the oscillator
equation
$a^+a \,\xi_1(x) = \varepsilon\, \xi_1(x)$ together with 
$\xi_2 = (a^+)^k \xi_1$, provided $\xi_1 \propto a^k \xi_2$, so
that the nonphysical eigenfunction has the form
$\Xi=(c_1 \xi_1, c_2 \xi_2)$. The coefficients $c_1,c_2$  
depend now on $\varepsilon$ and can be easily computed, but, in
general, they are complex too.

\section{Darboux transformation of the \jc Hamiltonian\label{DT-JC}}

To construct the Darboux transformation for the \jc Hamiltonian we
are using the existing approach developed for the matrix \Sc
equation \cite{GV,SP}. Therefore, we
first reduce the \jc eigenvalue problem to the \Sc
equation with a matrix-valued potential.
Then applying Darboux transformations we get new exactly solvable
matrix potentials for the \Sc Hamiltonian and
the use of the inverse transformation allows us to obtain
desired transformation for the \jc Hamiltonian.

 We start with
rewriting $h_{JC}$ in the coordinate representation as
\be
h_0=-\p_x^2+V_0+b\g \p_x \,,\quad
\g= \left(\begin{array}{cc} 0 & 1 \\ -1 &0
\end{array}\right),\quad
h_{JC}=\textstyle{\frac 12}h_0
\ee
with
\be
V_0=\left(
\begin{array}{cc} x^2-1+2\a &  bx \\ bx & x^2-1
\end{array}
\right)\,, \quad b=\sqrt 2\b\, ,
\ee
where, for simplicity, we have taken $\b$ real. The operator $h_0$
contains the first order derivative term, therefore the method of
\cite{GV,SP} cannot be applied to it directly. This undesirable term
is easily removed by the unitary transformation
\be\label{UT}
\wt h_0=\bU^+h_0\bU\,, \quad \bU=(\bU^+)^{-1}=e^{\frac 12 b\g x}
\ee
which gives us the \Sc Hamiltonian
\be
\wt h_0=-\p_x^2+\wt V_0\,,\quad
\wt V_0=\bU^+V_0\bU-\textstyle{\frac 14}\,b^2\,.
\ee

The Darboux transformation operator $\wt L$
for a \Sc type Hamiltonian is known to be
in the form \cite{GV,SP}
\be\label{tL}
\wt L=\p_x-\wt W\,,\quad \wt W=\ctU_x\ctU^{-1}\, ,
\ee
leading to the transformed potential
\be\label{V1t}
\wt V_1=\wt V_0-2\wt W_x\,.
\ee
Here and in the following the subscript $x$
means the derivative with respect to $x$
and the
tilde over a symbol marks a quantity related with the
\Sc equation, unless otherwise indicated.
The matrix $\ctU$ is a solution to the matrix eigenvalue problem
 \be\label{SEut}
\wt h_0\ctU =\ctU\L
\ee
with eigenvalue $\L$  being a diagonal matrix,
$\L =\rm{diag}(\l_1,\l_2)$.
In this case $\l_{k}$ are eigenvalues of
$\wt h_0$ corresponding to column-vectors
$\widetilde U_k=(\wt u_{1,k},\wt u_{2,k})^t$, $k=1,2$,
and the matrix $\ctU$ is just
composed of these columns, $\ctU=(\widetilde U_1,\widetilde U_2)$.

Once the Hamiltonian $\wt h_1=-\p_x^2+\wt V_1$ is determined we
realize the inverse transform to express all quantities in terms
of the eigenvalue problem for $h_0$. For instance, $\cU =\bU\, \ctU$
will be an $h_0$ matrix eigensolution, preserving the same eigenvalue
$\Lambda$,
\be
h_0\cU =\cU\L \,.
\ee
As for the potential
$V_1$, defined by the expression 
$h_1=\bU\wt h_1\bU^+:=-\p_x^2+V_1$, we get
\be\label{V1}
V_1=V_0-2W_x+b(\g W-W\g)\,.
\ee
It is easy to see that the operator $L$ (\ref{tL})
is covariant under the transformation $\Bbb U$:
\be \label{L}
L=\bU\wt L\bU^+=\p_x-W\,.
\ee
Keeping the usual SUSY terminology (see e.g. \cite{msusy})
we will refer to the matrix $W:=\cU_x\cU^{-1}$
as a {\it \jc superpotential}.
It is interesting to notice that the change in the transformed
potential (\ref{V1}) has two
contributions.
The term $-2W_x$
in (\ref{V1})
 corresponds to the
usual expression for the potential difference
due to the Darboux transformation (see e.g. \cite{MS,BS-rev}).
The next term proceeds from the first derivative member of the
Hamiltonian $h_0$ and coincides with the potential difference for
Darboux transformed Dirac systems (see \cite{NPS}).

We see from (\ref{V1}) and (\ref{L}) that the new potential and
the transformation operator are defined in terms of a matrix-valued
 solution, $\cU$, of the initial Hamiltonian. Therefore, as usual
 \cite{BS-rev} we call it {\it transformation
 function}. To keep the potential difference regular we have to
 impose an additional condition on $\cU$: $\mbox{det}\,\cU\ne 0$.
 Another additional condition must be imposed on $\cU$ if we
require that
$V_1$
 be represented by a self-adjoint matrix.
It is not difficult to see that if we restrict ourselves to real
potentials, the last term in (\ref{V1}) is symmetric for any
matrix $W$. Therefore, the potential $V_1$ will be self-adjoint
if $W$ is real and its derivative symmetric, i.e.,
 $W-W^+=C_{JC}$ where $C_{JC}=-C_{JC}^+$ is a skew Hermitian constant
 which because of the real character of $W$ should be proportional to
 $\g$, $C_{JC}=c_0\g$.
Now using the relation
$W=\bU \wt W\bU^+ + \frac b2\g$ we find
\be\label{ss}
\wt W-\wt W^+=c\g \,,\quad c=c_0- b\,.
\ee
We notice that  both superpotentials $W$ and
$\wt W$ can never be simultaneously Hermitian. 
In section \ref{partners} we illustrate that both possibilities (the
hermiticity of
$W$ or that of $\wt W$) may take place.
From here it follows that the usual approach
of the multichannel supersymmetric quantum mechanics
based on Hermitian superpotential matrices,
$\wt W=\wt W^+$,
cannot by applied to the Hamiltonian (\ref{UT}). The above
condition should be replaced by a weaker demand
(\ref{ss}).

So, we have established
how the matrix Darboux transformation should be modified
when applied to the \jc Hamiltonian.
 It can be considered as a special case of the
general transformation introduced by Goncharenko and Veselov
\cite{GV} with an additional condition of type (\ref{ss}).
Now we are going to investigate
implications this restriction imposes in general and then we
will apply obtained results to the particular case of the
\jc Hamiltonian.

\section{Properties of the Darboux transformation for
the matrix Schr\"odinger equation}

To make easier the notation in this section
we will suppress all tildes over the
quantities related with the \Sc equation introduced in the previous
section.

Darboux transformation of the usual (one-component) \Sc equation has
a number of nice properties (see e.g. \cite{BS-rev}) making it popular
in different fields of theoretical and mathematical physics. In this
section we show how some of them are translated to the matrix level.
In our opinion the most interesting properties are the following:

\begin{itemize}
\item
The most convenient way to introduce the transformation operator
associated with the name Darboux is to define it as a {\it differential
intertwiner} between two Hamiltonians $h_0$ and $h_1$.
Defining (formally)
an adjoint operation such that the Hamiltonians are
self-adjoint one gets the adjoint intertwining relation
\be\label{inter}
Lh_0=h_1L\,,\quad L^+h_1=h_0L^+\,.
\ee

\item
Given $L$ and $L^+$ one can compose them thus discovering {\it
factorization properties} first discussed by \Sc and studied in
detail by Infeld and Hull.

\item
In matrix notation
the above properties can equivalently be rewritten as commutation and
anticommutation relations giving rise to a {\it supersymmetry
algebra} commonly known also as {\it supersymmetric quantum mechanics}.

\item
There exists a procedure allowing one to realize a {\it one-to-one
correspondence} between the spaces of solutions of the
Hamiltonians $h_0$ and $h_1$. Therefore, if $h_0$ is exactly
solvable, then $h_1$ is solvable as well. In this case it is
rather easy to find the spectrum of $h_1$ when the operators are
defined in a Hilbert space. In particular, one can conclude that
the spectrum of
$h_1$ coincides with the spectrum of $h_0$ with the possible exception
of a finite number of levels. In the simplest case either the spectra
may differ only by the ground state level (so called case of the
``exact supersymmetry") or they may be identical (``broken
supersymmetry").

\item
Making use of a special two-step differential transformation one recovers
 an {\it integral transformation}
(sometimes called Luban-Pursey method)
 leading, in general,
 to a relation between the method of Darboux transformations and
 inverse scattering method \cite{JP95}.

\item
There exist nice Crum-Krein formulae for a compact representation
of the resulting action of a chain of first order transformations.
Recently they have been generalized to the matrix case \cite{SP},
so we will not discuss this point here.

\end{itemize}

\subsection{Factorization properties and underlying supersymmetry}

First we will study how the factorization properties
are translated to the matrix level.

Since
\be\label{Wd}
W^+=W-C\,,\quad W=\cU_x\cU^{-1}
\ee
\be\label{SE}
h_0\cU =\cU\L
\ee
we have
$L^+=-\p_x-W+C$ which results in the following superposition
\be\label{LdL}
L^+L=h_0+g_0\,,\quad g_0=CL-\cU\L\cU^{-1}\,.
\ee
To find the superposition $LL^+$ we express $L$ in terms of $W^+$:
$L=\p_x-W^+-C$ which yields
\be\label{LLd}
LL^+=h_1+g_1\,,\quad g_1= -CL^+-(\cU\L\cU^{-1})^+\,.
\ee
From here it follows that $g_0=g_0^+$ commutes with $h_0$ and
$g_1=g_1^+$ commutes with $h_1$ and they are intertwined by $L$,
$Lg_0=g_1L$. The facts which can also be checked by the
direct calculation.
Nevertheless, it is necessary to stress here that
this observation is correct only formally.
When all operators are considered as acting in a Hilbert space
they may be not commuting because of different domains of
definitions and the subject should be studied more carefully.

From (\ref{SEut}) and its adjoint form
it follows another nice property of the transformation function
$\cU$:
\[
2(\cU\L\cU^{-1})_x=CV_1-V_0C=Ch_1-h_0C
\]
which shows that for the particular case $C=0$ the matrix
$\cU\L\cU^{-1}$ is constant.
We would like also to stress that even in this particular case
 the factorizations (\ref{LdL}) and (\ref{LLd})
do not coincide with the ones giving rise to the usual
multichannel supersymmetric quantum mechanics:
$L^+L=h_0-\lambda I$ and $LL^+=h_1-\lambda I$
where $\lambda$ is known as factorization constant
and $I$ is the unit matrix. This fact was
already mentioned in \cite{SP}.
Another point worth to be  noticed is that in contrast with the paper
\cite{Andr}, where similar symmetry operators have been found, we give
their explicit expression in terms of a solution $\cU$
 of the initial eigenvalue problem. In fact
 such nontrivial symmetries may exist only if the Hamiltonian $h_0$
 has at least one matrix solution $\cU$ with the property
(\ref{Wd}).

 Another interesting observation is that if the spectrum of $h_0$
 is nondegenerate and
 essentially self adjoint operators $h_0$ and $g_0$ have a
 common set (dense in the Hilbert space) in their domains of
 definition,
 where they commute,
 they
 should be related by a functional dependence. This statement
 follows from the fact that if the spectrum of $h_0$ is
 nondegenerate then a complete set of operators commuting with $h_0$
 consists of only one operator which is $h_0$ itself.
 Hence, by definition of a complete set of operators
 (see, e.g \cite{Ber}) any other self adjoint operator commuting with
 $h_0$ is a function of $h_0$.
In this case  all eigenfunctions of $h_0$ can be found by
 solving the eigenvalue problem for $g_0$,  which is a first order
 differential operator. We shall see latter that just the
 \jc Hamiltonian presents a nontrivial example of such
 a situation.

From the point of view of supersymmetric quantum mechanics we can
construct here a wider (extended) superalgebra
than the one usually appearing in the scalar case.
In addition to the
superhamiltonian
\[
H=\left(
\begin{array}{cc}
h_0 & 0 \\0 &h_1
\end{array}
\right)
\]
and mutually conjugated nilpotent supercharges
\[
Q=(Q^+)^+=\left(
\begin{array}{cc}
0 & 0 \\L &0
\end{array}
\right)
\]
we have the symmetry operator
\[
G=\left(
\begin{array}{cc}
g_0 & 0 \\0 &g_1
\end{array}
\right).
\]
Now, as usual, the intertwining relations (\ref{inter}) are equivalent to the
commutation of $H$ and $G$ with the supercharges and the
factorization properties are translated to the anticommutation
relation for the supercharges $QQ^++Q^+Q=H+G$. It is also evident
that $H$ and $G$ commute between them.
According to the remark made in the previous paragraph, these
relations are also rather formal and when a concrete Hilbert space
is considered it is necessary to take care of domains
where the operators act.

In the case of $2\times 2$ uncoupled \Sc equations with $V_0=0$ the
 conjugation property for the superpotential can readily be
 analyzed. We have found that the condition (\ref{Wd})
 can be satisfied only for $C=0$ meaning that differential first
 order symmetries do not exist in this case.
 On the other hand, the result of
 the previous section shows that matrix Hamiltonians with such
 symmetry operators do exist.
Another useful remark concerns the usual multichannel
supersymmetry approach. It imposes the restriction on $\L$ to be
diagonal with equal elements. In our case $\L$ is allowed to have
different diagonal elements which enlarges considerably the set of
exactly solvable partners for a given $V_0$.

\subsection{One-to-one correspondence between the spaces of
solutions}

Being differential, the operators $L$ and $L^+$ have nontrival kernels.
Therefore, the correspondence between the spaces of solutions of
the \Sc equations for $h_0$ and $h_1$ should be studied carefully.
More precisely, a detailed analysis is necessary for kernel
spaces of these operators which are eigenfunctions of $h_0$ and
$h_1$ corresponding to the eigenvalue matrix $\L$.
In the scalar case $\mbox{Ker}(h_{0}-\l )$ is a two dimensional
space easily determined by a solution $u$, $h_0u=\l u$.
This possibility is based on the property of the
Wronskian of two solutions corresponding to the same eigenvalue to
be equal to one.
First we are going to show how this property translates to the matrix
case and then how it helps us to establish a one-to one
correspondence.

The conjugation equation for the superpotential (\ref{Wd})
being rewritten in another form
\be\label{W}
W(\cU ,\cU):=\cU^+_x\cU-\cU^+\cU_x+\cU^+C\cU=0
\ee
establishes a property of the transformation function $\cU$.
We relate it with the fact that the Wronskian of a
solution with itself vanishes.
Therefore we call the
quantity $W(\cU ,\cU)$, defined by the middle part of (\ref{W}),
Wronskian with coinciding arguments.
Quite naturally
for different arguments it is defined as:
\be\label{mW}
W(\cV ,\cU):=\cV^+_x\cU-\cV^+\cU_x+\cV^+C\cU \,.
\ee
We notice that for $C=0$ and $\cU$, $\cV$ being usual functions
(not matrices) it coincides with the usual expression for the
Wronskian. Therefore, we consider (\ref{mW}) as a matrix generalization
of Wronskian.

Now we will show that if $\cU$ is a solution to (\ref{SE}), satisfying
(\ref{Wd}),
then there exists another solution to the same equation, $\cV$,
which can be found from the condition
$W(\cV ,\cU)=-1$.

We proceed first to find an explicit expression for $\cV$.
Multiplying the equation $W(\cV ,\cU)=-1$ by $\cU^{-1}$ from the
right and using (\ref{Wd}) we get
\be\label{vdx}
\cV^+_x-\cV^+(\cU^+)^{-1}\cU^+_x=\cU^{-1}\,.
\ee
Multiplying the adjoint form
of (\ref{vdx})
by $\cU^{-1}$ from the left we obtain
$(\cU^{-1}\cV)_x=\cU^{-1}(\cU^+)^{-1}$, which gives us the final
result
\be\label{V}
\cV=\cU\int_{x_0}^x(\cU^+\cU)^{-1}dx + \cU C_1 \,.
\ee
In general,
we cannot neglect the matrix integration constant $C_1$. This is
due to the
evident property of (\ref{SE}) not to be a linear equation with
respect to the multiplication on matrices.
Only if this
constant takes a definite value this function is a solution to
(\ref{SE}).

Now we proceed to find $C_1$.
For this purpose we put the function (\ref{V}) into equation
(\ref{SE}). Using the same equation for $\cU$ and (\ref{Wd}) we
obtain
\be\hspace{-3em}\label{r1}
V_0\cU C_1- C_1(\cU^+)^{-1}-\cU_{xx}C_1=
u\int_{x_0}^x(\cU^+\cU)^{-1}dx\L-u\L\int_{x_0}^x(\cU^+\cU)^{-1}dx
+\cU C_1\L \,.
\ee
Here it is necessary to calculate the commutator of $\L$ with the integral at the
right hand side of (\ref{r1}) but first we have to find the
commutator of $\L$ with $(\cU^+\cU)^{-1}$.

From  (\ref{Wd}) and (\ref{SE}) it is easy to get
\[
\L\cU^+\cU-\cU^+\cU\L=(\cU^+C\cU)_x
\]
which can readily be transformed to
\[
\L(\cU^+\cU )^{-1}-(\cU^+\cU)^{-1}\L =(\cU^{-1}C(\cU^+)^{-1})_x
\]
which gives us
\[
\int_{x_0}^x(\cU^+\cU)^{-1}dx\,\L-\L\int_{x_0}^x(\cU^+\cU)^{-1}dx=
-\cU^{-1}C(\cU^+)^{-1}+C_2\,.
\]
Here
\be\label{C2}
C_2=(\cU^{-1}C(\cU^+)^{-1})_{x=x_0}=-C_2^+\,.
\ee
Now once again using (\ref{SE})
we obtain from (\ref{r1}) the equation for $C_1$:
\be\label{C1}
C_2+C_1\L -\L C_1=0\,,\quad C_1^+=C_1\,.
\ee
 Since we just reduced the \Sc equation for
$\cV$ to the last equation for $C_1$, this means that with the
constant $C_1$ thus determined the function $\cV$ (\ref{V}) satisfies the
\Sc equation (\ref{SE}).

Here we would like to notice  that for any $n$-dimensional vector space
Ker\,$(h_0-E)$
of
solutions of the eigenvalue equation for $h_0$ with a given $E\ne
\l_k$, $k=1,2,\ldots ,n$, $h_0\psi_E=E\psi_E$,
(we remind that $\l_k$ correspond to the eigenvectors $U_k$ of the matrix
$\cU$) the kernel of the operator $L$ is the empty set whereas
$L\cU =0$.
To find solutions of the \Sc equation with the
eigenvalues $\l_k$ we can act by $L$ on the function
$\cV$ which yields
\be\label{tcv}
\cuU=L\cV =(\cU^+)^{-1}\,,\quad h_1\cuU=\cuU\L\,.
\ee
Since $\L$ is supposed to be diagonal, columns of $\cuU$ are
eigenvectors of $h_1$ with eigenvalues $\l_k$.

An obvious but necessary remark is that $L^+\cuU=0$.
Moreover, it is easy to see that
equation (\ref{Wd})
 is covariant under the
transformation $\cU\to \cuU = (\cU {\,}^+)^{-1}$ meaning that $\cuU$
 has the
same property. Hence, another solution of the eigenvalue problem
for $h_1$ corresponding to the matrix eigenvalue $\L$ can be found
using the same formula (\ref{V}) applied this time to $\cuU$:
\be\label{vt}
\cvV =(\cU^+)^{-1}\left(\int_{x_0}^x\cU^+\cU dx +
\overline C_1 \right) \,,\quad
h_1\cvV=\cvV\L \,.
\ee
Once again because of the diagonal character of $\L$ columns
of $\cvV$ are eigenvectors of $h_1$ with eigenvalues $\l_k$.
The constant $\overline C_1$ from (\ref{vt}) should be determined in a similar way
that $C_1$ from (\ref{V}).

So, for the eigenvectors $\vfi_E$ of $h_1$,
$h_1\vfi_E=E\vfi_E$,
 with $E\ne \l_k$, $k=1,\ldots ,n$ the kernel of
$L^+$ is the empty set,
Ker\,$L^+=\emptyset$.
This means that if
$\psi_E$ is also an eigensolution to $L^+L\psi_E=\mu_E\psi_E$
then $\vfi_E=L\psi_E$ satisfies the equation
$LL^+\vfi_E=\mu_E\vfi_E$.
In this case
$L$ and $L^+$ realize a one-to-one correspondence between the
spaces of solutions for all $E\ne \l_k$, $k=1,\ldots ,n$.
For $E = \l_k$ this correspondence may be
established by hand $\cU\to \cuU$,  $\cV\to \cvV$
and continued by linearity at the level of vector-valued
eigenfunctions $\psi_E$ and $\vfi_E$ of $h_0$ and $h_1$.
If  $L^+L$
 get $\psi_E$ beyond the one-dimensional space Span$\psi_E$,
$L^+L\psi_E\notin \mbox{Span}\psi_E$,
 this
correspondence is more subtle and needs an additional analysis.

\subsection{Integral transformations}

Let a matrix constant $C_{01}$ be such that $(\cU^+)^{-1} C_{01}$ be
the solution to the equation (\ref{tcv}). Then the function
$\wt\cV_1=\cvV +(\cU^+)^{-1} C_{01}$
with $\cvV$ as given in (\ref{vt})
is the solution to the same equation and it can be taken as the
transformation function for the next transformation step. After
some algebra one gets
\be\label{W2}
(\wt\cV_1)_x(\wt\cV_1)^{-1}=
-W^+ +W_2\,,\quad
W_2:=\cU \left[\int_{x_0}^x\cU^+\cU dx +C_0\right]^{-1}\cU^+
\ee
where $C_0=C_{01}+\overline C_1$.
So, the potential $V_2=V_1-2(\wt\cV_1)_x(\wt\cV_1)^{-1}$ is given
by
\be\label{Vint}
V_2=V_0+\Delta V_2\,,\quad
\Delta V_2= -2(W_2)_x\,.
\ee
The potential difference $\Delta V_2$
from (\ref{Vint}) and (\ref{W2})
is self-adjoint for any matrix $\cU$.
Nevertheless, we
can use here only these $\cU$s which give rise to a self-adjoint
derivative of the superpotential, $W_x^+=W_x$, since only under
the condition (\ref{Wd})
the function (\ref{vt}) is a solution to the \Sc
equation after the first transformation step.
An advantage of this
formula with respect to (\ref{V1t})
could be a much easier possibility to get an everywhere
nonsingular resulting potential since just the constant $C_0$ can be
used for this purpose whereas in (\ref{V1t}) we have not any
freedom of this kind.
Moreover, in such a way one can get families of isospectral and
isophase (known also as phase-equivalent) potentials.
Applying twice appropriately changed formula (\ref{tL})
we express solutions of the \Sc equation with the potential
(\ref{Vint}) in terms of solutions of the initial eigenvalue
problem
\be\label{xi}
\vfi_E =\cU\L\cU^{-1}\psi_E -\psi_E E-
(\,C+W_2 ) (\psi_{Ex}-\cU_x\cU^{-1}\psi_E)
\ee
\[
h_2\vfi_E =\vfi_E E\,,\quad h_2=-\p_x^2 +V_2\,.
\]
Here $\psi$ may be both vector-valued and matrix-valued eigenfunction
of $h_0$ with $E$ being a number in the first case and a diagonal
matrix in the second case, $h_0\psi_E =\psi_E E$. As usual, for
$E=\L$ and $\psi_E=\cU$, the right hand side of (\ref{xi})
vanishes,
 but the missing eigen-solutions correspond to $({\wt\cV_1}^{\,+})^{-1}$.

\section{SUSY partners for the \jc Hamiltonian\label{partners}}

\subsection{Properties of the Darboux transformation for the
\jc Hamiltonian}

Using the relation between the \jc Hamiltonian
and the \Sc Hamiltonian given by equation (\ref{UT}) we can readily
apply the results of the previous section to the \jc
eigenvalue problem. From now on we restore notations of the section
\ref{DT-JC} and will distinguish all quantities related with the
matrix \Sc equation by putting tilde over. The quantities without
the tilde will be used for the \jc problem.

First we notice that according to (\ref{ss})
even for a self-adjoint \jc superpotential
the constant $C$ introduced in the previous section
is not zero but $C=b\g$.
Factorization properties (\ref{LdL}) and (\ref{LLd})
together with the symmetry operators are covariant under this
transformation. Therefore the symmetry operator for the \jc
Hamiltonian can be found from a diagonal superpotential matrix.
Since $\Bbb U$ commutes with $C=b\g$ the formula
(\ref{C2}) for the constant $C_2$ is covariant also.
The unitary transformation  (\ref{UT}) does not affect the
eigenvalue matrix $\L$. Therefore the equation (\ref{C1}) for the constant
$C_1$ remains intact.
The equations (\ref{tcv}) and (\ref{vt}) for solutions of the
transformed equations with the eigenvalue $\L$ are also covariant
but now they are valid for the self-adjoint \jc superpotential
$W$.
For the integral transformation the transformed potential is given
by the same formula (\ref{V1}), where $W$ is replaced by
$W_2$, which is given by (\ref{W2}), where all quantities are
related now with the \jc system. The formula (\ref{xi}) for solutions
of the equation with this potential is also covariant.

\subsection{Examples\label{examples}}

{\bf 1.}
In the simplest case the transformation function can be taken in
the form (see section \ref{JC})
\be
\cU =\left(
\begin{array}{cc}
0  & e^{x^2/2} \\ e^{-x^2/2} & 0
\end{array}
\right).
\ee
It corresponds to the diagonal eigenvalue matrix $\L
=\mbox{diag}(0,2\alpha -2)$ and diagonal superpotential
$W=\mbox{diag}(x,-x)$. Formula (\ref{V1}) gives us the transformed
potential
\be
V_1=\left(
\begin{array}{cc}
x^2+2\a -3  & -b x \\ -b x & x^2+1
\end{array}
\right)
\ee
and finally the transformed Hamiltonian
\be
h_1=
2\left(
\begin{array}{cc}
a^+a+\a -2  &  -\b a^+ \\ -\b a & a^+a
\end{array}
\right).
\ee
Solutions of the \Sc equation with this Hamiltonian can be
obtained from the solutions of the \jc eigenvalue
problem using the transformation operator
\be
L=
\sqrt 2 \left(
\begin{array}{cc}
a^+  &  0 \\0 & a
\end{array}
\right).
\ee
Using (\ref{LdL}) we find the symmetry operator
\be\label{g00}
g_0=g_{00}+4\,,\quad
g_{00}=b\g \p_x +
\left(
\begin{array}{cc}
2\a -2   &  bx \\bx & 0
\end{array}
\right)= 2
\left(
\begin{array}{cc}
\a -1   &  \b a \\\b a^+ & 0
\end{array}
\right).
\ee
The spectrum of the \jc Hamiltonian is nondegenerate
and the set of all finite linear combinations of its
eigenfunctions is dense in the corresponding Hilbert space. The
operator $g_{00}$ (\ref{g00})
 is well-defined on this set where it commutes with
$h_{JC}$. Hence,  as expected
 the latter is a second order polynomial of
$g_{00}$:
$b^2h_0=g_{00}(g_{00}+b^2-2\a +2)$.

{\bf 2.}
Another possibility is to construct the transformation function
$\cU$ from two vectors of the type $(c_1\psi_n ,c_2\psi_{n+1})$
with the coefficients $c_{1,2}$
given in (\ref{c1})
corresponding to the energies $E_n^{1,2}$ (\ref{en})
at $k=1$. For $n=0$ we easily get the following superpotential
\be
W=\left(\matrix{ -x & 0 \cr 0 & \frac{1}{x} - x \cr}\right).
\ee
We notice that it is singular at the origin and
cannot give rise to a self-adjoint potential for the spectral problem
on the whole real line.
Nevertheless, up to a constant shifting it produces the same symmetry
operator (\ref{g00}).
Moreover, since at any $n$ both
even and odd Hermit polynomials are involved in this construction,
the superpotentials we can obtain in this way
are singular at the origin for any $n$.

{\bf 3.}
As it usually happens in the method of Darboux transformations (see e.g. \cite{BS-rev}) a
singular point can be removed by the next transformation if it is
realized with the help of a transformation function corresponding
to the adjacent spectral point. Moreover, two
consecutive first order transformations can be replaced by a
single second order transformation (for details see \cite{SP}).
As a result we get the following potential difference
\[
\Delta V= \frac{4}{(1 + 2\,x^2)^2}
\left(
\matrix{ (1 + 2\,x^2)^2 & b\,x(1 + 2\,x^2) \cr
b\,x(1 + 2\,x^2)& 4\,x^4 + 8\,x^2 -1)\cr  }
     \right).
\]
We notice here that the spectrum of the Hamiltonian $h_2=h_0+\Delta V$
coincides with the spectrum of $h_0$ except for four levels
 $E_{0,1}^{1,2}$ which are removed by this transformations.
 It is also clear that taking other values of $n$ one can remove
 any four levels $E_{n,n+1}^{1,2}$.
 For this purpose we are using  two transformation functions
 \[
\cU_1=
\left(
\begin{array}{rr}
\psi_n(x) & -A_1\psi_n(x)\\
A_1\psi_{n+1}(x)&\psi_{n+1}(x)
\end{array}
\right),\quad
\cU_2=
\left(
\begin{array}{rr}
\psi_{n+1}(x) & -A_2\psi_{n+1}(x)\\
A_2\psi_{n+2}(x)&\psi_{n+2}(x)
\end{array}
\right)
\]
and the algorithm developed in \cite{SP}.
After some algebra we find that the second
order superpotential is diagonal,
$W_2=\mbox{diag}(w_0,w_1)$, with the entries
$w_{k}=2\psi_{n+k}\psi_{n+k+1}/
(\psi'_{n+k}\psi_{n+k+1}-\psi_{n+k}\psi'_{n+k+1})$,
$k=0,1$.
Here in the denominator we see the Wronskian of two consecutive
discrete spectrum eigenfunctions of the usual Harmonic oscillator
potential which is known to be nodeless \cite{BS-rev}. Therefore
it gives ``good" potential differences for any non-negative
integer $n$.
 For instance, $n=1$ case
 corresponds to the following potential difference
\[
\Delta V= 4
\left(
\matrix{ \frac{4x^4+8x^2-1}{(1+2x^2)^2}
& b\left( \frac{4x^2}{3+4x^4}-\frac{x}{1+2x^2} \right)
\cr
 b\left( \frac{4x^2}{3+4x^4}-\frac{x}{1+2x^2} \right)&
1+\frac{8x^2(4x^4-9)}{(3+4x^4)^2}
\cr  }
     \right).
\]

{\bf 4.}
The creation of new energy levels is also possible. For this
purpose we need solutions of the \Sc equation, which do not belong
to the Hilbert space discussed in section \ref{JC}.
They are constructed with the help of the functions
$\vfi_n(x)=\psi_n(ix)$.
For the transformation function
\[
\cU=
\left(
\begin{array}{cc}
\vfi_0 &
A\vfi_{n+1}\\
0  &
\vfi_n
\end{array}
\right),\quad n=0,1,\ldots
\]
the superpotential reads
\[
W=
\left(
\begin{array}{cc}
\frac{\vfi'_0(x)}{\vfi_0(x)} &
A\frac{\vfi_0(x)\vfi'_{n+1}(x)-\vfi'_0(x)\vfi_{n+1}(x)}%
{\vfi_0(x)\vfi_n(x)}\\[0.3em]
 0 &
\frac{\vfi'_n(x)}{\vfi_n(x)}
\end{array}
\right).
\]
Using the recurrence relation for the Hermit polynomial we find that
its non-zero off-diagonal element does not depend on $x$ thus
giving us a nontrivial example of
a non-Hermitian up to a constant superpotential.
Substituting the logarithmic derivatives by their expressions we
get its final form
\[
W=
\left(
\begin{array}{cc}
x &
A\sqrt{2n+2}\\[0.3em]
 0 &
x+\frac{2niH_{n-1}(ix)}{H_n(ix)}
\end{array}
\right).
\]
We notice that for even $n$ the Hermit polynomials $H_n(ix)$ are
nodeless whereas $H_{n-1}(ix)$ are purely imaginary. So, for
$n=0,2,\ldots$ this formula gives us ``good" potential
differences provided $A$ is real which is not always the case.
For instance, for $n=0$ the potential difference is a
constant diagonal matrix
$\Delta V=\mbox{diag}(B_0,-4 -B_0)$,
where $B_0=-1-\a \pm \sqrt{(\a-1)^2-2b^2}$.
Two possible signs here and below are related with different signs in
(\ref{en}).
The first nontrivial case corresponds to $n=2$:
\[
\Delta V=
\left(
\begin{array}{cc}
B_1&
\frac{4bx}{1+2x^2}\\[0.3em]
\frac{4bx}{1+2x^2}&
-4\,\frac{4x^4+3}{(1+2x^2)^2}-B_1
\end{array}
\right)
\]
where $B_1=-1-\a \pm \sqrt{(\a-1)^2-6b^2}$. We have to notice that after
being shifted by a constant diagonal matrix this potential
difference may be reduced to the previous example at $n=0$.
Hence, to get an essentially new potential difference
we have to consider $n=4$:
\[
\Delta V=
\left(
\begin{array}{cc}
B_2&
\frac{8bx(3+2x^2)}{4x^4+12x^2+3}\\[0.3em]
\frac{8bx(3+2x^2)}{4x^4+12x^2+3}&
-4\,\frac{16x^8+64x^6+120x^4+45}{(4x^4+12x^2+3)^2}-B_2
\end{array}
\right)
\]
where $B_2=-1-\a  \pm\sqrt{(\a-1)^2-10b^2}$.

{\bf 5}.
As the final example we give a one-parameter family of potentials
isospectral with \jc Hamiltonian obtained with the help of the
integral transformation (\ref{Vint}), (\ref{W2}):
\[
\Delta V= 8
\left(
\begin{array}{cc}
\frac{1+xF(x)}{F^2(x)}&
-\frac b4(\frac{1}{F(x)}+\frac{2x^2}{2x-F(x)})\\
-\frac b4(\frac{1}{F(x)}+\frac{2x^2}{2x-F(x)})&
2x\frac{2x+(x^2-1)F(x)}{(2x-F(x))^2}
\end{array}
\right)
\]
where $F(x)=\sqrt\pi e^{x^2}(c+\mbox{erf}(x))$ and real $c$ is
such that $|c|>1$.

\section{Conclusion}

A careful analysis of the matrix Darboux transformation method
has
permitted us to establish
 such properties
as: ({\it i\/}) a new factorization scheme,
which is responsible on appearance of an extended supersymmetry
underlying matrix Hamiltonians
and hidden symmetry operators, ({\it ii\/}) a
one-to-one correspondence between the spaces of
solutions, which permits to readily determine the changes in the
spectrum, and ({\it iii\/}) the integral transformation formula, which
gives families of isospectral Hamiltonians.
We applied these results to get a new symmetry operator for the \jc
Hamiltonian and build up many of its exactly solvable partners.
A future step
of this research will be
the investigation of
physical phenomena, such as for instance the collapse and revival
for new Hamiltonians.

\section*{Acknowledgment}

The work is partially supported by the Spanish
Ministerio de Education, Cultura y Deporte Grant SAB2000-0240 and
the Spanish MCYT (projects BFM2002-02000 and
BFM2002-03773), Junta de Castilla y Le\'on (VA085/02), and European
FEDER grant. 
The work of BSF is partially supported by President Grant of Russia
1743.2003.2. The authors are grateful to L~M~Nieto and M~Gadella for
helpful discussions.


\end{document}